\documentclass[a4paper,fleqn,usenatbib]{mnras}

\usepackage[T1]{fontenc}
\usepackage{ae,aecompl}

\usepackage{graphicx}   
\usepackage{amsmath}    
\usepackage{amssymb}    

\usepackage{txfonts}
\usepackage{hyperref}
\usepackage{breakurl}

\title[The polluted atmosphere of the white dwarf NLTT~19868]{Extreme abundance 
ratios in the polluted atmosphere of the cool white dwarf NLTT~19868\thanks{Based on observations collected at the European Organisation
for Astronomical Research in the Southern Hemisphere, Chile under programme
IDs 082.D-0750 and 093.D-0797.}}

\author[A. Kawka and S. Vennes]{Adela Kawka\thanks{E-mail: kawka@asu.cas.cz} and St\'ephane Vennes
\\
Astronomick\'y \'ustav AV \v{C}R, Fri\v{c}ova 298,CZ-251 65 Ond\v{r}ejov,
Czech Republic}

\date{Accepted XXX. Received YYY; in original form ZZZ}

\pubyear{2015}

\begin{document}
\label{firstpage}
\pagerange{\pageref{firstpage}--\pageref{lastpage}}
\maketitle

\begin{abstract}
We present an analysis of intermediate-dispersion spectra and photometric data 
of the newly identified cool, polluted white dwarf NLTT~19868. The spectra 
obtained with X-shooter on the Very Large Telescope (VLT)-Melipal show strong 
lines of calcium, and several lines of magnesium, aluminium and iron. We 
use these spectra and the optical-to-near infrared spectral energy 
distribution to constrain the atmospheric parameters of NLTT~19868. 
Our analysis shows that NLTT~19868 is iron poor with respect to aluminium and 
calcium. A comparison with other cool, polluted white dwarfs shows that the Fe 
to Ca abundance ratio (Fe/Ca) varies by up to approximately two orders of 
magnitudes over a narrow temperature range with NLTT~19868 at one extremum in 
the Fe/Ca ratio and, in contrast, NLTT~888 at the other extremum. 
The sample shows evidence of extreme diversity in the composition of the 
accreted material: In the case of NLTT~888, the inferred composition of 
the accreted matter is akin to iron-rich planetary core composition, while in 
the case of NLTT~19868 it is close to mantle or bulk-Earth composition depleted 
by subsequent chemical separation at the bottom of the convection zone. 
\end{abstract}


\section{Introduction}

Polluted white dwarfs (typed with a suffix ``Z'') provide an opportunity to 
investigate the ultimate fate of planetary systems. Although planets have not 
yet been detected around white dwarfs, the evidence for the presence of 
planetary debris around these objects lies in their polluted atmospheres. 
Approximately one quarter of white dwarfs show the presence of elements heavier 
than helium in their atmospheres \citep{zuc2003,zuc2010} and approximately
one fifth of these have a mid-infrared (IR) excess that is consistent
with a circumstellar, debris disc \citep{far2009}. More recently
using the Cosmic Origins Spectrograph on the \emph{Hubble Space Telescope}
\citet{koe2014} have shown that about half of DA white dwarfs with effective
temperatures ranging from 17\,000 K to 27\,000 K have polluted atmospheres.

Several detailed studies of polluted white dwarfs have uncovered large 
variations in the composition of the accreted material. Based on a study of 
ultraviolet (UV) spectra of a sample of white dwarfs 
(19\,000 < $T_{\rm eff} < 24\,000$ K), \citet{gan2012} showed that the 
abundance diversity in the accreted material is similar to that observed among 
Solar System meteorites, although the effect of selective radiative radiation 
pressure on accretion rate calculations was neglected. \citet{cha2014} 
demonstrated that selective radiation pressure on trace elements, silicon for 
instance, shapes observed abundance patterns in hot white dwarfs 
($\gtrsim 17\,000$ K). After including this effect in their calculations,
\citet{koe2014} concluded that at least 27\% of their white dwarf sample, which includes the
\citet{gan2012} sample, would be currently
accreting, while in 29\% of these objects, usually among the warmest in their sample, the effect of radiative levitation dominates
the abundance pattern. The inclusion of this effect also leads to a reduction
in the estimated accretion flow in some objects with $T_{\rm eff} > 20\,000$ K (e.g., WD0431+126).
An analysis of UV and optical spectra of two additional 
white dwarfs by \citet{xu2014} show the accreting source to be of a rocky 
nature where the abundance of refractory elements is enhanced compared to 
volatile elements. Also, \citet{zuc2011} showed that the cool, hydrogen-rich 
and magnetic white dwarf NLTT~43806 (typed DAZH) is enriched in aluminium but 
poor in iron which suggests that the accreting material is similar to the Earth 
lithosphere. Oxygen has been detected in several white dwarfs 
\citep[e.g., GALEX~J1931+0117; ][]{ven2010}, and, in some of these objects, 
the amount of oxygen with respect to the other heavier elements detected suggests that
the accreted material contains significant amount of water. For example, in
GD~61 \citet{far2013} found that the accreted material contains oxygen in
excess of the amount expected to be carried by metal oxides, suggesting that the
parent material contained water. A similar finding, but with a higher fraction
of water, was found in the case of SDSS~J124231.07$+$522626.6 \citep{rad2015}.

The material accreted at the surface of a white dwarf is subjected to diffusion processes:
trace elements are quickly mixed in the convective envelope of cool white dwarfs, and
diffuse-out below the convection zone in a period of time much shorter than evolutionary timescales \citep{paq1986}.
Recent estimates \citep{koe2006,koe2009} \footnote{See also 
\burl{http://www1.astrophysik.uni-kiel.de/~koester/astrophysics/astrophysics.html}}
of diffusion timescales show that relics of an accretion event remain visible in the
photosphere of a cool (6\,000~K) hydrogen-rich white dwarf for nearly $10^5$ years and much longer (several $10^6$ years) for cool helium-rich white
dwarfs. However, the observed 
abundance would follow details of the accretion history, and the presence of heavy elements
is likely transitory when compared to the cooling age of old white dwarfs ($\ga 10^9$ years).

We present a spectroscopic and photometric analysis of an hitherto unknown 
cool, polluted white dwarf (NLTT~19868) from the 
revised NLTT catalogue of \citet{sal2003}.
We provide details of the new observations in Section 2: We obtained new low-
and high-dispersion spectra as well as new and archival photometric
measurements allowing to build a spectral energy distribution (SED). 
In Section 3, we analyse our new data and derive atmospheric parameters: temperature, surface
gravity, and composition. Next, in Section 4, we attempt to reconstruct recent accretion history onto this
object. Then, we draw a comparison with the sample of
cool white dwarfs highlighting the peculiar photospheric composition of the cool
white dwarf NLTT~19868, and, finally, we summarize our results.

\section{Observations}

We present detailed spectroscopic and photometric observations of the newly identified white dwarf
NLTT~19868.

\subsection{Spectroscopy}

We first observed NLTT~19868 with the ESO Faint Object Spectrograph and
Camera (EFOSC2) attached to the New Technology Telescope (NTT) at La Silla
Observatory on UT 2009 March 3. Using grism number 11 (300 lines/mm) with 
the slit-width set to 1 arcsec, we obtained a resolution of 
$\Delta \lambda \approx 14$ \AA. The two consecutive spectra of 1800 s each 
revealed a cool DAZ white dwarf with strong \ion{Ca}{ii} H\&K lines.

We followed up on our initial observations with four sets of echelle spectra of 
using the X-shooter spectrograph \citep{ver2011} attached to the UT3 at 
Paranal Observatory on UT 2014 May 1, 29 and June 1. The slit-width was set to 
0.5, 0.9 and 0.6 arcsec for the UVB, VIS and NIR arms, respectively. This setup 
provided a resolving power of 9900, 7450 and 7780 for the UVB, VIS and 
NIR arms, respectively. The exposure times for the UVB and VIS arms were 2940 
and 3000 s, respectively, and for the NIR arm we obtained five exposures of 
600 s each.

\subsection{Photometry}

We used the acquisition images from the EFOSC2 and X-shooter observations
to obtain estimates of $R$ and $V$ magnitudes of NLTT~19868, respectively.
First, we measured the instrumental magnitudes of NLTT~19868 and of a brighter
comparison star (RA[J2000]=08h 36m 03.44s, Dec[J2000]=$-$10\degr 05\arcmin 52\farcs 5) with 
published photometry ($V=14.974\pm0.023$ mag,\ $g=15.252\pm0.021$ mag,\ 
$r=14.872\pm0.052$ mag, and $i=14.729\pm0.121$ mag) from the AAVSO Photometric 
All-Sky Survey (APASS) \footnote{\url{http://www.aavso.org/apass}}.
APASS is an all-sky survey conducted in five filters (Johnson $B$ and $V$, and
Sloan $g$, $r$ and $i$) with a magnitude range from approximately 10 up to 17.
We converted the Sloan $r$ magnitude to the Johnson $R$ magnitude using the 
transformation equations of Lupton (2005)
\footnote{\burl{http://www.sdss.org/dr12/algorithms/sdssUBVRITransform/\#Lupton2005}}:
\begin{equation}
R = r - 0.1837(g - r) - 0.0971,
\end{equation}
or
\begin{equation}
R = r - 0.2936(r - i) - 0.1439.
\end{equation}
We calculated $R$ using both equations and used the average of the two measurements
($R=14.70$ mag). Finally, using the difference between the instrumental magnitudes
of NLTT~19868 and the comparison star, we calculated $V=17.55\pm0.02$ mag and
$R=17.16\pm0.04$ mag for NLTT~19868. Note that the uncertainties for $V$ and $R$
are statistical only and neglect any possible systematic effects.

We obtained IR photometric measurements from the Two Micron All Sky Survey
\citep[2MASS; ][]{skr2006} and {\emph Wide-field Infrared Survey Explorer} 
\citep[\emph{WISE}; ][]{wri2010}. The measurements which are all on the 
Vega system are listed in Table~\ref{tbl_phot}. The
$W3$ and $W4$ measurements are not listed because only upper limits were
available for this object. We have examined the \emph{WISE} images in combination
with our X-shooter $V$ acquisition images. These images show that there
is a nearby star 5.2 arcsec away at a position angle P.A. = 98$^\circ$. 
The crowded star does not share the white dwarf proper motion and, consequently, is
not physically related. Using the proper
motion of NLTT~19868
the distance between the nearby star and NLTT~19868 would 
have been 5.2 arcsec at P.A. = 107$^\circ$ at the time the \emph{WISE} images were obtained
(2010). Since the point spread function (PSF) of $W1$ and $W2$ are 6.1 and 6.4
arcsec, respectively, the \emph{WISE} photometric measurements listed in Table~\ref{tbl_phot}
should combine both NLTT~19868 and the nearby object. The amount of contamination is
unknown since the spectral type of the nearby star is unknown, although its
SED shows it to be a cool object ($T\approx 4\,400$\,K).

We have examined the 2MASS images which show that NLTT~19868 and the nearby 
object are clearly separated, and we conclude that the 2MASS photometric 
measurements of NLTT~19868 are not contaminated by the nearby star. The 
2MASS catalog did not contain any flags that would suggest problems with the
$JHK$ photometry.

\begin{table}
\centering
\caption{Photometry and astrometry}
\label{tbl_phot}
\begin{tabular}{lcc}
\hline
Parameter & Measurement & Reference \\
\hline
R.A.(J2000)   &    08h 36m 01.65s & 1 \\
Dec (J2000)   & $-$10\degr 06\arcmin 07\farcs 31 & 1 \\
$\mu_\alpha$  & $-0.0182\pm0.0055$ \arcsec yr$^{-1}$ & 1 \\
$\mu_\delta$  & $-0.2214\pm0.0055$ \arcsec yr$^{-1}$ & 1 \\
$V$           & $17.55\pm0.02$ mag & 2 \\
$R$           & $17.16\pm0.04$ mag & 2 \\
$J$           & $16.025\pm0.073$ mag & 3 \\
$H$           & $15.765\pm0.102$ mag & 3 \\
$K$           & $15.808\pm0.255$ mag & 3 \\
$W1$          & $14.675\pm0.035$ mag & 4 \\
$W2$          & $14.970\pm0.098$ mag & 4 \\
\hline
\end{tabular}\\
References: (1) \citet{sal2003}; (2) This work; 
(3) 2MASS \citep{skr2006}; (4) \emph{WISE} \citep{wri2010}
\end{table}

\section{Analysis}

For our analysis of NLTT~19868, we used a grid of hydrogen-rich model
atmospheres calculated in local thermodynamic equilibrium. These models are
described in \citet{kaw2006} and \citet{kaw2012}. We include bulk heavy
elements at fixed abundance in the charge neutrality equation along with identifiable
elements (Table~\ref{tbl_line}) at a variable abundance. The bulk composition was held fixed at 
${\rm [Z/H]}\equiv \log{\rm Z/H}-\log{\rm Z/H}_\odot=-4.0$ and includes
abundant elements from carbon to zinc. The solar abundance is defined by
\citet{asp2009}.
Strong resonance lines of calcium are also included in the radiative equilibrium
equation resulting in a marked drop in the surface temperature at $\tau_{\rm Rosseland}<<1$.

\subsection{Stellar parameters}

\begin{figure}
\includegraphics[width=\columnwidth]{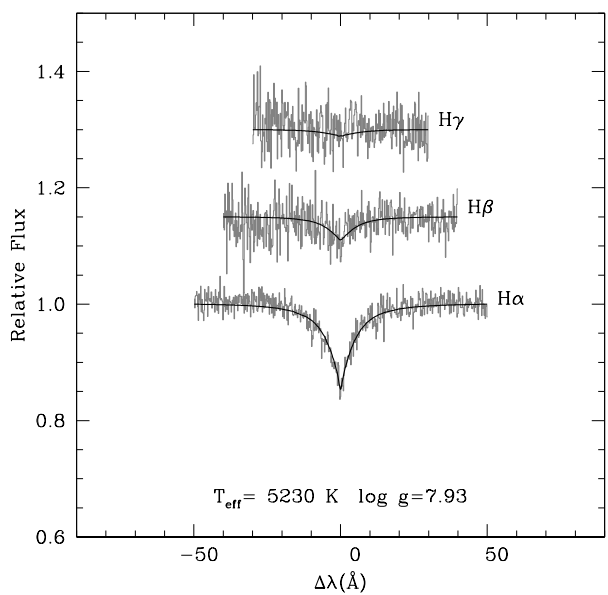}
\caption{Balmer line profiles of NLTT~19868 compared to the best fitting model
spectrum.}
\label{fig_prof}
\end{figure}

We determined the effective temperature and surface gravity of NLTT~19868 by 
fitting the Balmer lines in the X-shooter spectra with synthetic spectra using $\chi^2$ minimization
techniques. Fig.~\ref{fig_prof} shows the observed H$\alpha$, H$\beta$ and H$\gamma$ 
lines compared to the best-fitting model spectrum at $T_{\rm eff}=5\,230\pm100$ K
and $\log{g}=7.93\pm0.24$.
Given the cool nature of NLTT~19868 only H$\alpha$ and H$\beta$ are detected,
while H$\gamma$ is essentially levelled. This causes large uncertainties in the $\log{g}$ measurement.

\begin{figure*}
\includegraphics[viewport=1 155 560 575, clip,width=0.9\textwidth]{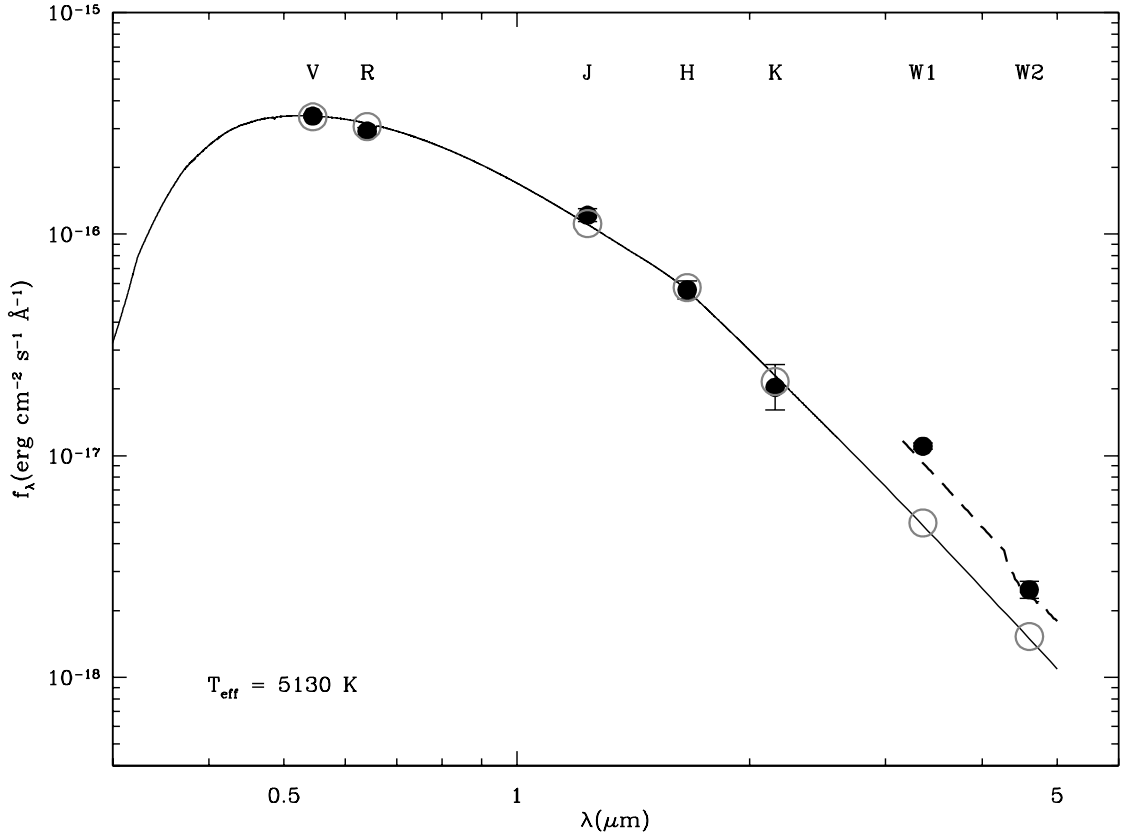}
\caption{The observed SED (full circles), based on the available photometric 
data (Table~\ref{tbl_phot}), compared the best-fitting synthetic SED 
(open circles). For comparison, the best-fitting model at $T=5\,130$ K and
$\log{g} = 7.9$ is also shown (full line). The \emph{WISE} bands are contaminated
by a cool, crowded star: The dashed line shows the 3 to 5$\mu$m segment of the
K5~V template normalized to the optical bands.}
\label{fig_sed}
\end{figure*}

We also calculated the effective temperature by fitting the photometric
magnitudes with those calculated from model spectra. For reasons given above (Section 2.2),
we excluded
the \emph{WISE} measurements from the analysis. Fig.~\ref{fig_sed} shows the observed SED compared
to the best-fitting model at $T_{\rm eff}=5\,130$ K and $\log{g}=7.9$. 
The effective temperatures from the spectroscopic fit and the SED
fit are consistent. 

To confirm our initial suspicion that the
\emph{WISE} photometric measurements of NLTT~19868 are contaminated, we have also
attempted to model the crowded star given the limited amount of data. Following the
same method adopted for the white dwarf, we calculated $V=17.98\pm0.02$ and $R=17.24\pm0.02$ mag 
using the X-shooter and EFOSC2 acquisition images and the APASS. Then, we
attempted to determine the spectral type using Kurucz templates fitted to the available $V$, $R$, 2MASS
$J=15.569\pm0.059$, $H=15.070\pm0.057$, $K=14.763\pm0.103$ mag and \emph{WISE} 
photometric measurements. Assuming that the star is on the main-sequence
our best estimate of the spectral type corresponds to K5~V. Fig.~\ref{fig_sed}
compares the expected IR flux from the white dwarf
to the contaminated $W1$ and $W2$ photometric measurements as well as our estimates of the
contribution from the nearby cool star. We conclude that most of the 3 to 5$\mu m$ flux from the
crowded star and that the object is probably a K5~V star.

Based on the 2MASS measurements, we also conclude that the SED of NLTT~19868 
does not show an IR excess. This behaviour is consistent with that of 
other cool DAZ and DZ white dwarfs
($T_{\rm eff} < 7\,000$ K), with none of them showing an IR excess in 
\emph{Spitzer} \citep{deb2007,far2008,far2009}, 2MASS, or \emph{WISE} 
\citep[see ][]{xu2012,jur2013} observations. Recently,
\citet{ber2014} have shown that the fraction of polluted white dwarfs
with a debris disc revealed by IR excess decreases with cooling age 
with an apparent downturn above 600 Myr, or $T_{\rm eff}\lesssim10\,000$K.
This suggests that either the brightness of debris discs 
decreases with age or that the source of material dries out with time.
The coolest known white dwarf to show IR excess due to the presence of
a debris disc is the DAZ G166-58 with an effective temperature of 7400 K 
\citep{far2008}. The cool DZ G245-58 with an effective temperature of 7500 K
also appears to show some IR excess, however it remains to be confirmed.

For the rest of this work, we have adopted the weighted average of the
effective temperatures from the spectral and SED fits. The adopted surface
gravity was updated to match the adopted effective temperature.
We calculated the mass and cooling age using the evolutionary mass-radius
relations of \citet{ben1999}.

\begin{table}
\begin{minipage}{\columnwidth}
\centering
\caption{Main spectral line identifications}
\renewcommand{\footnoterule}{\vspace*{-15pt}}
\renewcommand{\thefootnote}{\alph{footnote}}
\label{tbl_line}
\begin{tabular}{lcc}
\hline
Ion & $\lambda$ (\AA) \footnotemark[1]\footnotetext[1]{Wavelengths from \burl{http://www.nist.gov/pml/data/asd.cfm} at the National Institute of Standards and Technology (NIST).} & E.W. (\AA) \\
\hline
\ion{Fe}{i} & 3581.193 & 0.34 \\
\ion{Fe}{i} & 3719.934 & 0.13 \\
\ion{Fe}{i} & 3734.864 & 0.19 \\
\ion{Fe}{i} & 3737.131 & 0.12 \\
\ion{Fe}{i} & 3745.561 & 0.11 \\
\ion{Fe}{i} & 3749.485 & 0.08 \\
\ion{Fe}{i} & 3758.233 & 0.08 \\
\ion{Fe}{i} & 3820.425 & 0.09 \\
\ion{Mg}{i} & 3832.304 & 0.13 \\
\ion{Mg}{i} & 3838.292 & 0.28 \\
\ion{Fe}{i} & 3859.911 & 0.13 \\
\ion{Ca}{ii}& 3933.663 & 4.31 \\
\ion{Al}{i} & 3944.006 & 0.51 \\
\ion{Al}{i} & 3961.520 & 0.40 \\
\ion{Ca}{ii}& 3968.469 & 2.93 \\
\ion{Ca}{i} & 4226.728 & 0.97 \\
\ion{H}{i}  & 4861.323 & 0.72 \\
\ion{H}{i}  & 6562.797 & 2.34 \\
\ion{Ca}{ii}& 8542.090 & 0.44 \\
\ion{Ca}{ii}& 8662.140 & 0.15 \\
\hline
\end{tabular}
\end{minipage}
\end{table}

We measured the radial velocity of NLTT~19868 in each of the four spectra using H$\alpha$ and the calcium
lines \ion{Ca}{i}$\lambda 4226$ and \ion{Ca}{ii} H\&K. Using H$\alpha$ we measured the 
barycentric corrected velocity $\varv = -16.4\pm9.9$ km~s$^{-1}$.
The average velocity measurement of the calcium lines in 
all four spectra results in a barycentric corrected velocity of 
$-22.2\pm4.1$ km~s$^{-1}$. The calcium and hydrogen measurements are consistent within 
uncertainties. However, since the calcium lines are much stronger and sharper,
the spread in their measurements is much lower than those of the H$\alpha$
measurements. Therefore, we adopted the radial velocity based on the three calcium lines,
$\varv=-22.2\pm4.1$ km~s$^{-1}$. Taking into account the expected gravitational 
redshift based on the adopted parameters, $\gamma_g=24.7\pm6.6$ km~s$^{-1}$, we
determined that the actual velocity of NLTT~19868 is $\varv_r = -46.9\pm7.8$ km~s$^{-1}$.

Adopting the algorithm of \citet{joh1987} and correcting for the Solar motion
using \citet{hog2005}, we calculated the Galactic velocity vectors 
$U,V,W = 56,20,-30$ km~s$^{-1}$. The kinematics of NLTT~19868 suggests that it 
belongs to the Galactic thin disc \citep[see ][]{sou2003}. Assuming
that NLTT~19868 is a thin disc star and that the total age
of the star does not exceed 10 Gyr \citep{liu2000,del2005}, then the lower 
limit on the white dwarf mass is $0.52\ M_\odot$ as determined from the
progenitor lifetime versus final mass relations for $Z=0.02$ of \citet{rom2015}.

NLTT~19868 does not appear to be magnetic with a surface-averaged field 
limit of $B_S < 40$ kG set by the X-shooter instrumental resolution.

\begin{table}
\centering
\caption{Properties of NLTT~19868}
\label{tbl_prop}
\begin{tabular}{lccc}
\hline
Parameter & Spec. & SED & Adopted \\
\hline
$T_{\rm eff}$ (K)  &  $5\,230\pm100$   & $5\,130\pm230$ & $5\,220\pm90$ \\
$\log{g}$ (cgs)    &  $7.93\pm0.24$    & ...          & $7.90\pm0.22$ \\
Mass ($M_\odot$)   &  $0.54^{+0.15}_{-0.12}$ & ...    & $0.52^{+0.13}_{-0.10}$  \\
Cooling Age (Gyr)  &  $3.6\pm1.0$      &  ...         & $3.4\pm0.9$ \\
Distance (pc)      &  $35\pm6$         &  ...         & $36\pm6$ \\
$\varv_r$ (km~s$^{-1}$)& $-48.2\pm8.5$     &  ...     & $-46.9\pm7.8$ \\
$U$,$V$,$W$ (km~s$^{-1}$)&  ...        &  ...         & 56,18,$-$30 \\
\hline
\end{tabular}
\end{table}

\subsection{Photospheric composition}

We computed detailed spectral syntheses using the adopted atmospheric 
parameters. The model line profiles include the effect of Stark and 
van der Waals broadening mechanisms following a procedure described in 
\citet{kaw2012}. We fitted the observed line profiles with a grid of spectra 
at fixed temperature and surface gravity but with varying abundances. The 
models were convolved with a Gaussian function set to the instrumental 
resolving power ($R\approx 9000$). We also tested the effect of temperature and 
surface gravity variations ($\Delta T=100$\,K and $\Delta\log{g}=0.2$\,dex).

\begin{figure*}
\includegraphics[width=0.99\columnwidth]{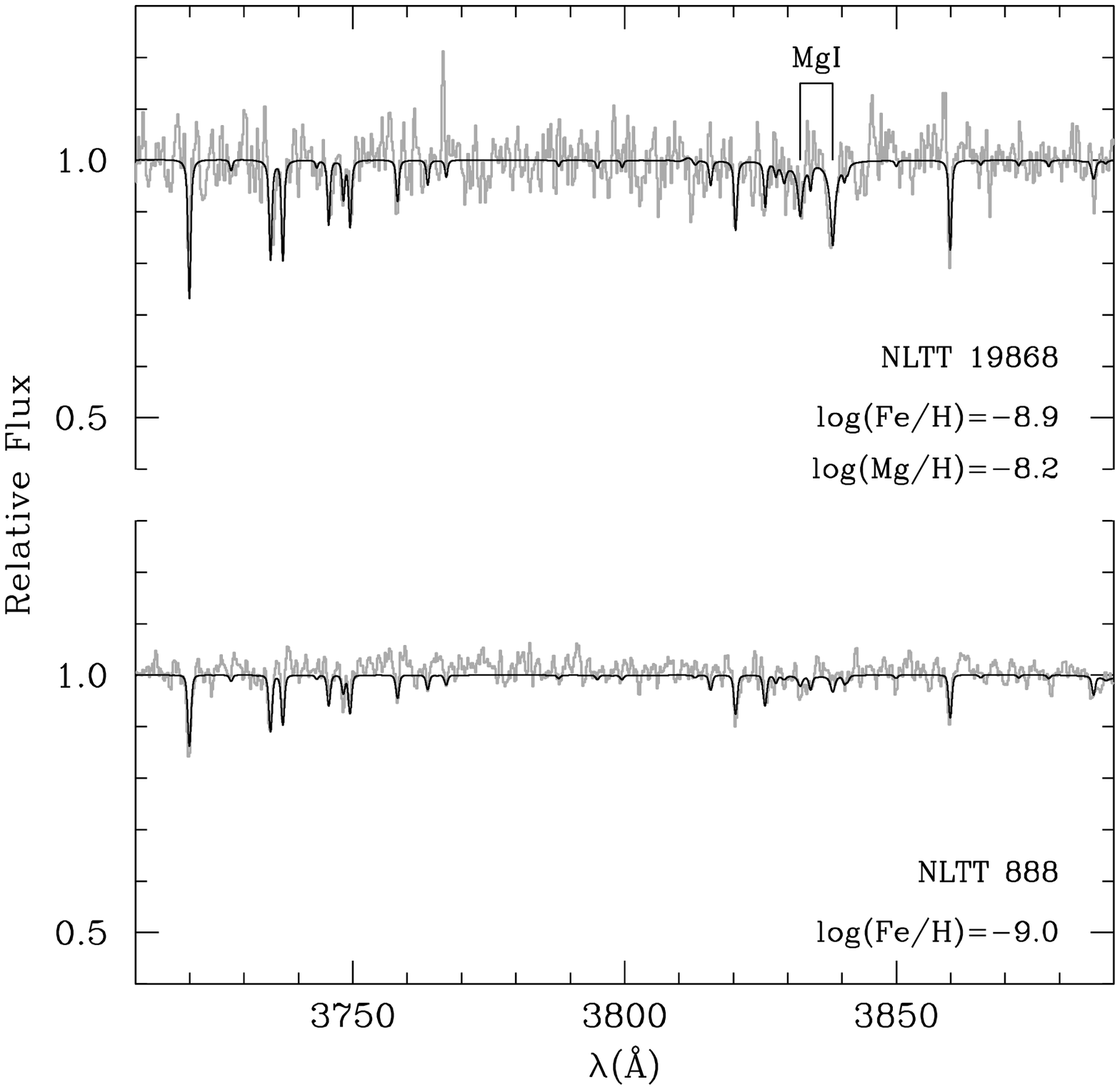}%
\includegraphics[width=0.99\columnwidth]{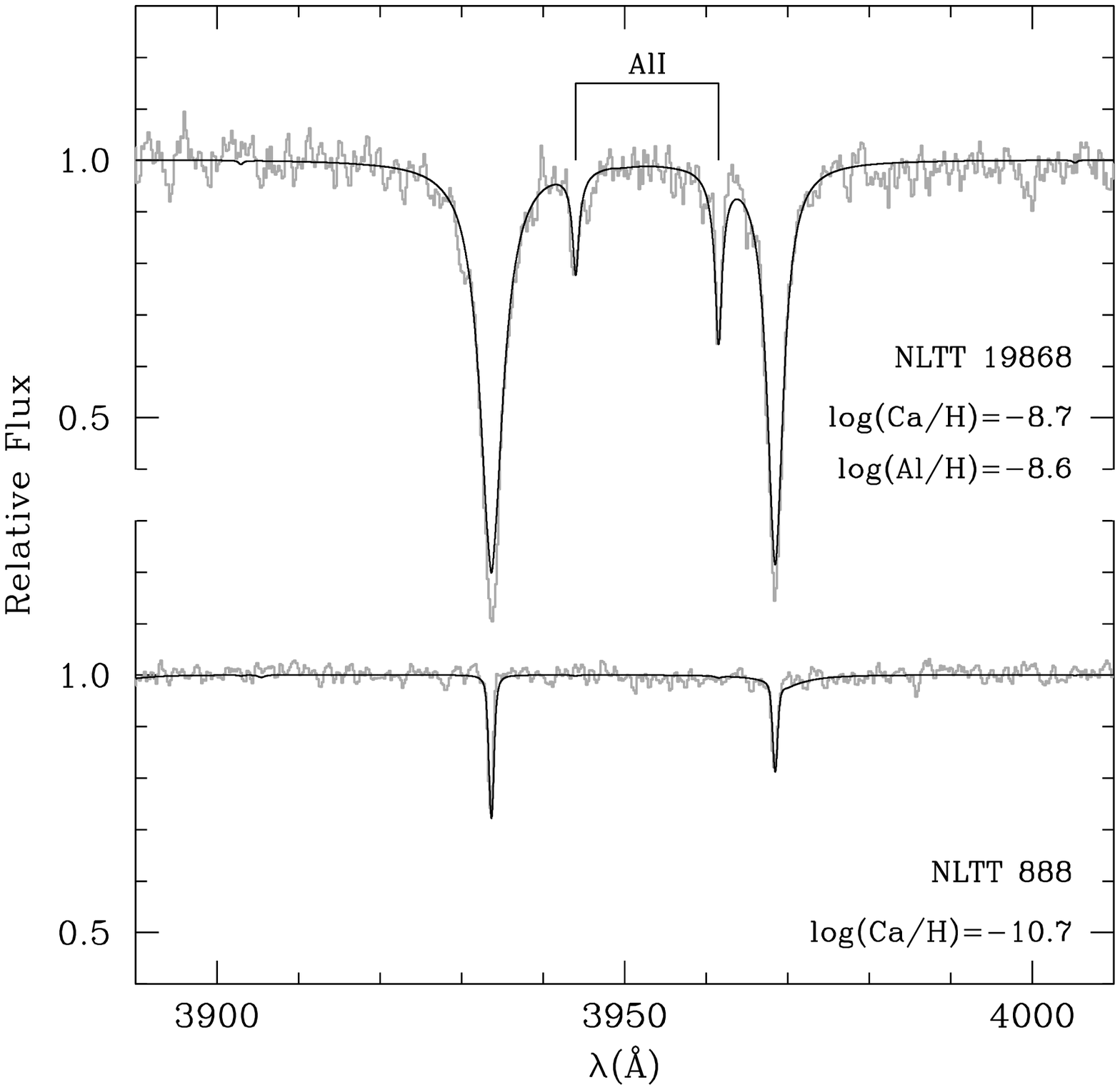}
\caption{X-shooter spectra of NLTT~19868 and NLTT~888 and best-fitting model
spectra. Lines of Mg, Al, and Ca are notably stronger in NLTT~19868
although the Fe lines are comparable in strength to those observed in NLTT~888.}
\label{fig_comp}
\end{figure*}

Figure~\ref{fig_comp} shows the X-shooter spectra of NLTT~19868 compared 
to the best-fitting model spectra. The resulting abundance pattern in NLTT~19868 shows marked contrast with that of the DAZ 
NLTT~888 \citep{kaw2014} which exposes a very high Fe/Ca ratio (see 
Section 4 for a detailed discussion). 
Table~\ref{tbl_abun} summarizes the abundance measurements. Magnesium is the 
most abundant element in the atmosphere of NLTT~19868, while iron is the least 
abundant element identified. However, when scaled to solar abundances, calcium 
and aluminium show the largest excess as compared to magnesium and iron. 
Interestingly, the Fe to Ca abundance ratio in NLTT~19868 is the lowest 
ever observed among polluted white dwarfs \citep[see ][]{jur2013}. 

Systematic effects due to temperature or surface gravity variations are generally small compared to statistical measurement errors.

\begin{table}
\centering
\begin{minipage}{\columnwidth}
\caption{Photospheric composition of NLTT~19868}
\renewcommand{\footnoterule}{\vspace*{-15pt}}
\renewcommand{\thefootnote}{\alph{footnote}}
\label{tbl_abun}
\begin{tabular}{lcccc}
\hline
  &                 &                                                                                        & \multicolumn{2}{c}{$\Delta\log{\rm Z/H}$} \\
                                                                                                               \cline{4-5} 
Z & $\log{\rm Z/H}$ & [Z$/$H] \footnotemark[1]\footnotetext[1]{[Z$/$H]$=\log{\rm Z/H}-\log{\rm Z/H}_\odot$.} & ($\Delta T=100$\,K) & ($\Delta\log{g}=0.2$\,dex) \\
\hline
Mg & $-8.20\pm0.30$ & $-3.80$ & $+0.00$ & $+0.10$ \\
Al & $-8.63\pm0.18$ & $-3.13$ & $+0.09$ & $+0.06$ \\
Ca & $-8.70\pm0.04$ & $-3.04$ & $+0.08$ & $+0.00$ \\
Fe & $-8.93\pm0.14$ & $-4.43$ & $+0.04$ & $+0.07$ \\
\hline
\end{tabular}
\end{minipage}
\end{table}

\section{Discussion and Summary}

Abundance studies of cool, hence old white dwarfs allow to determine the 
frequency of planetary debris at an advanced cooling age, i.e., long after the 
parent star left the main-sequence. Also, because of the longer diffusion 
timescales predicted in cool white dwarfs, diffusion effects may become more 
apparent, particularly following {a discrete accretion} event. In this 
case, a spread of diffusion timescales within a group of elements would lead 
to time-dependent alteration to abundance ratios allowing for a critical
examination of the physical conditions at the base of the convection zone \citep[see ][]{koe2009}. 

The number of cool DAZ white dwarfs remains low compared to the number of their
hot DAZ counterparts, or to the helium-rich DZ white dwarfs. 
Concentrating our efforts on the hydrogen-rich sample may help establish whether their
environment is similar to the more common DZs.
In this context, our analysis of the cool DAZ white dwarf NLTT~19868 and similar objects is 
timely. NLTT~19868 lies among the coolest known DAZ white dwarfs. Only three 
other DAZ white dwarfs have temperatures below 5500 K: G174-14 ($T=5\,139\pm82$ K, 
\citet{gia2012}), NLTT~10480 ($T=5\,200\pm200$ K, \citet{kaw2011}) and G77-50 
($T=5\,245\pm66$ K, \citet{far2011,gia2012}). The effective temperature adopted for
G77-50 is the weighted average of the two measurements from \citet{far2011} and
\citet{gia2012}. 

\begin{figure}
\includegraphics[width=1.0\columnwidth]{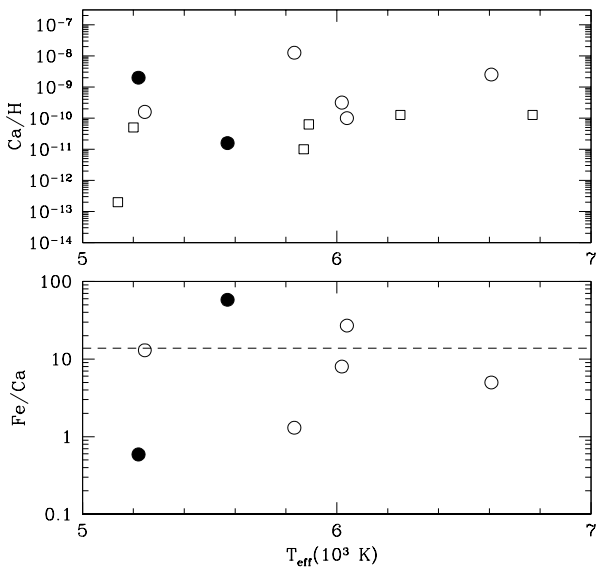}
\caption{Ca abundance (top) and the abundance ratio Fe/Ca (bottom) of all known
cool DAZ white dwarfs \citep[see ][ and references therein]{kaw2014} including 
NLTT~19868 from this work. The white dwarfs NLTT~19868 (left) and NLTT~888 
(right) are shown with full circles and cover the extrema in the Fe/Ca 
abundance ratio. The stars shown with open squares do not have a measured Fe abundance.
The Fe/Ca distribution among DAZ white dwarfs may be compared
to that of the DZ plus DAZ white dwarfs shown in \citet{jur2013}.}
\label{fig_ratio}
\end{figure}

Fig.~\ref{fig_ratio} (top panel) plots the calcium 
abundance of all known cool DAZ white dwarfs with temperatures lower than 
7\,000~K. The effective temperature for seven of these objects were updated 
with the results of \citet{gia2012}. The calcium abundance varies by several 
orders of magnitudes, consistent with other studies 
\citep[e.g., ][]{zuc2003,koe2005}. This range of abundances is possibly
a result of large variations in the accreted mass and of a possible time lapse
since the last accretion event resulting in a diffusion-induced decline in the 
observed abundance. Fig.~\ref{fig_ratio} (bottom panel) also shows the Fe/Ca 
abundance ratio for stars with a measurable iron abundance. Even though the 
sample is small, a large dispersion ($\sigma_{\log{\rm Fe/Ca}} \approx 0.7$) in the Fe/Ca abundance ratio is observed. 

Within the small sample of objects depicted in Fig.~\ref{fig_ratio}, 
$\log{\rm Fe/Ca}$ averages $\approx0.8$, which is lower than the 
bulk-Earth abundance ratio of $\approx1.1$ \citep{mcd2001,all2001} but
is still consistent given the $\sigma$ of the sample. We compared our 
cool sample to the larger sample of 50 polluted white dwarfs presented
by \citet{jur2013} which includes white dwarfs with 
$T_{\rm eff} \approx 5000$ K to $\approx 21\,000$ K. Our cool sample has a 
slightly smaller average but a larger dispersion than what is observed in the
\citet{jur2013} sample which has an average of $\log{\rm Fe/Ca}\approx1.0$ and 
a $\sigma \approx 0.4$. The likely reason for
the difference in the dispersion is that in the hotter sample the dispersion
is smaller than in the cooler sample and hence bringing down the dispersion in 
the \citet{jur2013} sample. 
Hence, 
the extrema at $\log{\rm Fe/Ca}\approx1.8$ \citep[NLTT~888; ][]{kaw2014}
and $\approx-0.2$ (NLTT~19868) are notable. The abundance ratio observed in the 
NLTT~888 is only second to that of the cool DZ white dwarf SDSS~J1043+3516 
\citep[$\log{\rm Fe/Ca} = 2.3$, ][]{koe2011} while Fe/Ca in NLTT~19868 is slightly lower than in the cool 
DAZ NLTT~43806 \citep[$\log{\rm Fe/Ca} = 0.1$, ][]{zuc2011}.

The observed abundance ratio $\log{\rm Fe/Al}=-0.3$ in NLTT~19868 is 
also the lowest known among known polluted white dwarfs, 
underlining the low relative abundance of Fe with respect to the other detected 
elements. This abundance ratio is slightly lower than that of NLTT~43806 
\citep[$\log{\rm Fe/Al}$=-0.2, ][]{zuc2011}.
Finally, the abundance ratio $\log{\rm Mg/Ca}=0.5$ is among the lowest in 
polluted white dwarf atmospheres, in fact it is the second lowest after the 
heavily polluted DBAZ GD~362 \citep[$\log{\rm Mg/Ca}=0.3$, ][]{zuc2007}.
A comparison with the population of the 60 polluted white dwarfs showing Mg 
\citep{jur2013}, for which the average is $\log{\rm Mg/Ca} = 1.26$ with a 
dispersion of 0.4, clearly places NLTT~19868 at the magnesium-deficient end of the distribution.

Diffusion at the bottom of the convection zone alters the observed abundance 
pattern over a time period much shorter than cooling timescales. Elements with 
short diffusion timescales relative to other elements are depleted faster and 
observed abundance ratios must vary over time. In a regime of steady state 
accretion, the hypothetical Z1/Z2 abundance ratio is simply given by
\begin{equation}
\frac{\rm Z1}{\rm Z2} = \Big{(}\frac{\rm Z1}{\rm Z2}_{\rm source}\Big{)} \times \frac{\rm \tau_{Z1}}{\rm \tau_{Z2}},
\end{equation}
while in the declining phase, i.e., after mass accretion ceased, the time ($t$) dependent abundance ratio
is given by
\begin{equation}
\log{\frac{\rm Z1}{\rm Z2}_{t}} = \log{\frac{\rm Z1}{\rm Z2}_{t=0}} -\frac{t}{\ln{10}} \Big{(} \frac{1}{\rm \tau_{Z1}}-\frac{1}{\rm \tau_{Z2}} \Big{)},
\end{equation}
where $\tau_{Z1,Z2}$ are the diffusion timescales at the bottom of the convection zone which
is assumed to be homogeneously mixed. Unfortunately, diffusion timescales are uncertain.
Few calculations are available for objects with effective temperatures below 6\,000~K.
The abundance pattern in the accreted material ${\rm Z1/Z2}_{\rm source}$ is also uncertain
and may, for example, correspond to solar \citep[see ][]{asp2009}, bulk, core, or mantle Earth
\citep{all2001,mcd2001}. Other types of material, based on an analogy with 
Solar System bodies such as those of meteorites or asteroids can also be
considered.

A study of the Fe/Ca abundance ratio in the extreme cases of NLTT~888 
($T_{\rm eff}\approx 5\,600$~K) and NLTT~19868 ($\approx 5\,200$~K) and of
applicable scenarios supports intrinsic abundance variations in the 
accreted material with likely alterations brought upon by diffusion 
effects. The observed Fe/Ca ratio is 58 in NLTT~888 and
0.59 in NLTT~19868. In both cases, steady-state accretion 
regime does not significantly alter the observed ratio relative to the 
parent material ratio. 

First, we examine the case of NLTT~888.
Interpolating the tables of \citet{koe2006} between 5\,000 and 6\,000~K, we 
find $\tau_{\rm Fe}/\tau_{\rm Ca} \approx 0.99$ and the abundance ratio in both 
media are nearly identical. Using timescales from \citet{paq1986}, 
$\tau_{\rm Fe}/\tau_{\rm Ca}\approx0.87$, leads to the same conclusion.
Therefore, the estimated ratio in the accreted material 
Fe/Ca $=50-60$ largely exceeds that of bulk- ($=$13.4) or mantle-Earth 
($=$1.8) suggesting that the accreted material in NLTT~888 consists of 
66\% core material and 34\% mantle using mass fractions converted to numbers
fractions for these various media
from \citet{mcd2001} and \citet{all2001}. 
Considering that the Fe/Ca abundance ratio is likely to decrease over a 
diffusion time scale ($4\times10^4$ years) in the eventuality that accretion is 
turned off, the inferred fraction of core material found in the parent 
body of the accreted material of NLTT~888 must be considered a lower limit. 
The predominance of core 
iron material implies the likely presence of the, yet undetected, core elements 
such as nickel and sulfur \citep{mcd2001}. A significant amount of silicon 
\citep[6\% by mass, ][]{mcd2001} is also predicted to be present in the core, 
although if both core and mantle are to be accreted, most of the silicon would 
come from the mantle.
Among the larger population of polluted white dwarfs, including higher
temperature stars, such as in the sample of \citet{gan2012} iron-enrichment
is also observed, for example, when compared to silicon, both PG~0843$+$516 and
PG~1015$+$161 accrete material where the Fe/Si ratio is comparable to the
core Earth. However, a comparison of the Fe/Ca ratio for stars in their sample,
GALEX~1931$+$0117 is the one with the highest Fe/Ca ratio \citep{ven2010} and is only slightly
lower than that of NLTT~888.

Scenarios for NLTT~19868 suggest that, on the contrary, core-type material is 
largely absent and that the accreted material is most likely composed of 
mantle-type material. Assuming steady-state accretion, the low iron content in 
the atmosphere implies a similar deficiency in the parent material, 
i.e., significantly lower than the Earth's mantle composition. Such a 
deficiency of iron in the parent material, i.e., below that of the Earth's 
mantle, may not be necessary. The observed Fe/Ca abundance ratio is well
reproduced if the accretion of Earth's mantle like material turned off
$\approx 10^6$ years ago, allowing to further reduce the Fe/Ca abundance ratio
in the convection zone to the observed level.
Note that diffusion timescales tabulated by 
\citet{paq1986} are systematically longer than those tabulated by 
\citet{koe2006} by up to a factor of three. The difference arises because the
convection zone reaches deeper in envelope models used by \citet{paq1986} where 
diffusion would operate in a denser medium. The depth of the convection
zone at low temperatures ($T_{\rm eff}\la 6\,000$~K) is not affected by the 
various treatments of the mixing length theory but, instead by different 
treatments of the equation-of-state and of the conductive opacity 
\citep{tas1990}. Interpolating the tables of \citet{koe2006} and 
\citet{paq1986} between 5\,000 and 6\,000~K, we find 
$\tau_{\rm Fe}/\tau_{\rm Ca} \approx 0.87$ and 0.96, respectively. Assuming 
mantle composition in the source, the original Fe/Ca ratio of 1.8 would 
be reduced to the observed ratio of 0.59 in $t\approx 9\times10^5$
years following \citet{koe2006} or a much longer time of $t\approx 9\times10^6$ 
years following \citet{paq1986}. A longer elapsed time implied by the 
calculations of \citet{paq1986} requires an unrealistically large accretion
event ($10^{-3}\,M_\odot$). Using the shorter elapsed time implied by the 
calculations of \citet{koe2006} results we can calculate an initial iron 
abundance of Fe/H $= 3.6\times 10^{-6}$ using:
\begin{equation}
\frac{\rm Fe}{\rm H}\bigg |_{t=0} = \frac{\rm Fe}{\rm H}\bigg |_t \times e^{\,t/\tau},
\end{equation}
where the elapsed time since the accretion event is $t=9.5\times 10^5$ years
and the diffusion timescale is $\tau = 1.1\times 10^5$ years.
Assuming that the mass of the convection zone is $M_{\rm cvz} = 0.5\,M_\odot \times q_{\rm cvz}$
and $\log{q_{\rm cvz}}\approx -6$ \citep{paq1986,tas1990}, then $M_{\rm cvz}\approx 10^{27}$~g,
and the total mass of iron accreted onto the white dwarf is $2\times10^{23}$~g, or, assuming an iron mass fraction of 6\%
in the mantle, the original accretion event would have weighed $3\times10^{24}$~g. This mass corresponds to less than
one thousandth of the mass of the Earth ($\approx 5\times10^{-4}\,M_\oplus$). 
The same diffusion scenario, but employing Earth bulk material characterized
by a higher Fe/Ca abundance ratio than in the mantle, would require a larger
settling time scale (elapsed time $\approx 3\times 10^6$ years) to match the
low, present-day Fe/Ca abundance ratio. Assuming an accretion event as large
as the convection zone itself, even calcium would have disappeared below the
detection limit ($\approx 10^{-13}$) after $\approx 3\times 10^6$ years.

In summary, we have identified a new cool, polluted white dwarf showing strong 
lines of calcium among weaker lines of magnesium, aluminium, and iron. Our 
model atmosphere analysis revealed the lowest iron to calcium abundance ratio 
of any cool polluted white dwarf.
Applying heavy element diffusion models, we found that the accretion event 
involving another peculiar DAZ white dwarf, NLTT~888, and that involving 
NLTT~19868 are clearly distinguishable. The material accreted into the surface 
of NLTT~888 is composed mainly of the iron-rich planetary core material, while 
the material accreted onto the surface of NLTT~19868 is more akin to Earth 
mantle material. In the case of NLTT~19868, the accretion event most likely
occurred several diffusion timescales ago. Although these scenarios appear 
reliable, details of the calculations rest upon uncertain diffusion timescale 
calculations. 

\section*{Acknowledgements}

A.K. and S.V. acknowledge support from the Grant Agency of the Czech Republic
(13-14581S and 15-15943S) and Ministry of Education, Youth and Sports (LG14013).
This work was also supported by the project RVO:67985815 in the Czech Republic.
This publication makes use of data products from the Wide-field Infrared Survey 
Explorer, which is a joint project of the University of California, Los 
Angeles, and the Jet Propulsion Laboratory/California Institute of Technology, 
funded by the National Aeronautics and Space Administration.
This publication makes use of data products from the Two Micron All Sky Survey,
which is a joint project of the University of Massachusetts and the Infrared 
Processing and Analysis Center/California Institute of Technology, funded by 
the National Aeronautics and Space Administration and the National Science 
Foundation.

\label{lastpage}

\end{document}